\documentclass[aps,pra,twocolumn,showpacs,superscriptaddress]{revtex4}
\usepackage{graphicx}
\usepackage{amsmath}
\usepackage{amssymb}

\input{epsf}
\begin{document}

\title{Illustration of universal relations for trapped
four-fermion system with arbitrary $s$-wave scattering length}

\author{D. Blume}
\affiliation{Department of Physics and Astronomy,
Washington State University,
  Pullman, Washington 99164-2814, USA}
\author{K.~M. Daily}
\affiliation{Department of Physics and Astronomy,
Washington State University,
  Pullman, Washington 99164-2814, USA}

\date{\today}

\begin{abstract}
A two-component four-fermion system with 
equal masses, interspecies $s$-wave scattering length $a_s$
and vanishing intraspecies interactions
under external spherically symmetric harmonic confinement is considered.
Using a correlated Gaussian basis set expansion approach, we determine 
the energies and various structural properties of the energetically
lowest-lying gas-like state throughout 
the crossover for various ranges of the
underlying two-body potential. Extrapolating to the zero-range limit,
our numerical results
show explicitly that the total energy, the trap energy
as well as certain aspects of the pair distribution
function and of the momentum distribution are related through
the so-called integrated contact intensity $I(a_s)$.
Furthermore, it is shown explicitly that the 
total energy and the trap energy
are related through a generalized virial theorem that accounts for a
non-zero range.
\end{abstract}

\pacs{}

\maketitle

\section{Introduction}
\label{sec_intro}
Dilute equal-mass two-component Fermi gases at unitarity, i.e.,
gases with infinitely 
large interspecies $s$-wave scattering length $a_s$ and vanishing
intraspecies scattering length, 
show unique universal 
properties~\cite{bake99,ohar02,ho04,chan04,astr04,chan04a,thom05,wern06,son06}.
Here, the term dilute means that $n r_0^3$ is
much smaller than 1, where
$r_0$ denotes the range of the underlying two-body potential
and $n$ the density of the system (for trapped gases, 
$n$ stands for the peak density).
Intuitively, the existence of universal
properties of the unitary two-component Fermi gas
can be understood by realizing that the infinitely large 
$s$-wave scattering length $a_s$ does not establish a meaningful length
scale for the system.
Correspondingly, the
only length scale 
of the infinitely strongly interacting system
is set by the density (for homogeneous systems) or the trapping 
potential (for inhomogeneous systems). 
The fact that the unitary two-component Fermi gas 
is characterized by a single length scale makes it,
in certain respects, similar to the non-interacting 
system. 
The above argument has been formalized 
within a hyperspherical framework~\cite{wern06}. 
It has been shown, e.g., that the
harmonically trapped two-component Fermi gas is characterized,
just as the non-interacting system, by ladders of states whose
energies are spaced by $2 \hbar \omega$~\cite{tan04,wern06}, where 
$\omega$ denotes the angular trapping frequency.
The $2 \hbar \omega$ spacing has been verified 
semi-analytically for the trapped two- 
and three-fermion systems~\cite{busc98,wern06a} 
and numerically for systems with up to six fermions~\cite{blum07}.
In addition, a number of other universal relations have been
established at 
unitarity~\cite{bake99,ohar02,ho04,chan04,astr04,chan04a,thom05,wern06,son06}.

More recently, the notion of universal relations has been extended 
to dilute equal-mass
two-component Fermi gases with finite interspecies scattering length
$a_s$ for which the two-body range $r_0$ is small compared to
$a_s$~\cite{tan08a,tan08b,tan08c,wern08,braa08,braa08a}. 
While it has been established early on
that the behaviors of these systems
through the crossover 
is governed by the $s$-wave scattering length $a_s$ 
alone (see Ref.~\cite{gior08} for a review)
and that the system is, in this sense, universal, only recently has
there been significant progress in relating seemingly disconnected
quantities such as the total energy, the expectation value
of the trapping potential, the pair distribution
function and the momentum distribution
through universal relations that
involve a quantity termed the integrated contact intensity 
$I(a_s)$~\cite{tan08a,tan08b,tan08c}. 
Physically, the integrated contact intensity measures the ``number of pairs per 
unit length''. It vanishes for vanishing $a_s$, indicating the absence
of pairs, and increases as the system enters the strongly-interacting and
eventually the weakly repulsively-interacting BEC regime, where the
pairs turn into actual molecules~\cite{grei03,zwie03}.

This paper explicitly
tests, to the best of our knowledge for the first time,
the equivalence of four definitions of the integrated contact intensity
$I(a_s)$~\cite{tan08a,tan08b,tan08c}.
In doing so, 
we explicitly determine the integrated
contact intensity for the energetically lowest-lying
gas-like state of the four-fermion system.
The spin-balanced
four-fermion system
possesses non-trivial correlations and can, at least in principle,
be realized by loading a degenerate Fermi gas 
consisting of dimers into
an optical lattice (see, e.g., Ref.~\cite{stof06}). Once a situation 
with one molecule
per site is realized, 
neighboring sites can be merged and molecules dissociated
by changing the $s$-wave scattering length.
This scheme would allow for the realization of an array of four-fermion
systems with adjustable $s$-wave scattering length $a_s$.
The integrated contact intensity determined in Sec.~\ref{sec_contactfourbody},
combined with the imaginary part of the two-body $s$-wave 
scattering~\cite{braa08},
readily leads to a prediction of the atom losses due to
inelastic two-body collisions, an experimentally observable quantity. 
Our results thus provide not only an illustration of the
equivalence of various definitions of $I(a_s)$ but also pave the way for making
quantitative predictions.

Section~\ref{sec_contactgeneral} 
reviews the definition of the integrated contact intensity $I(a_s)$ for a
trapped two-component Fermi gas interacting through a zero-range potential
with $s$-wave scattering length $a_s$.
Section~\ref{sec_contactfourbody} illustrates the equivalence of 
various definitions of the integrated contact intensity numerically.
To this end, we 
solve the Schr\"odinger equation for the four-fermion 
system as a function of the range
of the two-body potential and extrapolate our solutions 
to the zero-range
limit.
Section~\ref{sec_generalvirial} applies
the generalized virial theorem derived by Werner~\cite{wern08}
to our finite-range four-body energies.
Lastly, Sec.~\ref{sec_conclusion} concludes.

\section{Definition of the integrated contact intensity}
\label{sec_contactgeneral} 
This section reviews the definitions of the integrated contact intensity
$I(a_s)$ introduced by Tan~\cite{tan08a,tan08b,tan08c}.
For concreteness, we consider a two-component equal-mass Fermi gas 
consisting of $N$ atoms
under external spherically symmetric harmonic confinement
in which unlike fermions interact through a short-range two-body 
potential $V_{\mathrm{tb}}$.
For this system, the model Hamiltonian $H$ can be written as
\begin{eqnarray}
\label{eq_ham}
H= \sum_{j=1}^N \left[ -\frac{\hbar^2}{2m} \nabla^2_j + 
\frac{1}{2}m \omega^2 \vec{r}_j^2 \right] +
\sum_{j=1}^{N_{\uparrow}} \sum_{k=1}^{N_{\downarrow}}
V_{\mathrm{tb}}(r_{jk}),
\end{eqnarray}
where $m$ denotes the atom mass,
$\omega$ the angular trapping frequency,
$N_{\uparrow}$ 
the number of spin-up atoms,
 $N_{\downarrow}$ the number of spin-down atoms,
and $\vec{r}_j$ the position vector of the $j$th atom measured with
respect to the center of the trap.
The spherically symmetric two-body potential $V_{\mathrm{tb}}$ 
depends on the interparticle distance 
$r_{jk}$, where $r_{jk}=|\vec{r}_j-\vec{r}_k|$.
Throughout, we assume that 
$V_{\mathrm{tb}}$ is characterized by two length scales,
the $s$-wave scattering length $a_s$
and
the range $r_0$; this assumption implies, e.g.,
that the $p$-wave scattering length and other higher
partial wave scattering lengths are zero.
The lengths $a_s$ and $r_0$, together with the
harmonic 
oscillator length $a_{\mathrm{ho}}$, 
where $a_{\mathrm{ho}}=\sqrt{\hbar/(m\omega)}$,
constitute the relevant length scales of the Hamiltonian $H$.
Section~\ref{sec_contactfourbody} discusses our solutions to the stationary 
Schr\"odinger equation $H \psi = E \psi$, where the many-body wave function
$\psi$ depends on the coordinates of all particles, i.e.,
$\psi = \psi(\vec{r}_1,\cdots,\vec{r}_N)$.
Throughout, we indicate the dependence of the eigenenergy $E$ on 
the $s$-wave scattering length $a_s$ and the range $r_0$ of the underlying
two-body potential explicitly, i.e., we write $E=E(a_s,r_0)$.

The integrated contact intensity $I(a_s)$ introduced by 
Tan~\cite{tan08a,tan08b,tan08c} applies to 
the Hamiltonian given in Eq.~(\ref{eq_ham}) in the limit that
the two-body potential $V_{\mathrm{tb}}$ is characterized by
a vanishing range $r_0$.
Although the zero-range limit is, strictly speaking, not realized 
in nature, cold atom systems come arbitrarily close:
The range of the two-body potential between two alkali atoms
is set by the van der Waals tail, which---to leading order---falls off
as $C_6/r_{jk}^6$; typical values of the van der Waals length 
for alkali atoms are of the order of
100$a_0$~\cite{koeh06}. 
The zero-range limit is thus approximately realized 
if the absolute value of the $s$-wave scattering length $a_s$
and the harmonic oscillator length $a_{\mathrm{ho}}$
are much larger than the van der Waals length.
This scenario can be realized in cold atom experiments by 
choosing a sufficiently small trapping frequency $\omega$
and by tuning the $s$-wave scattering length 
through the application of
an external magnetic field
in
the vicinity of a Fano-Feshbach resonance.
In the zero-range limit, the two-body interactions 
impose
the Bethe-Peierls boundary condition whenever two unlike particles 
approach each other (see, e.g., Ref.~\cite{gior08}), 
making even the strongly-correlated
many-body problem amenable to analytical treatments.

The integrated contact intensity $I(a_s)$ can be defined through four distinct
relationships [see (i) through (iv) below]~\cite{tan08a,tan08b,tan08c}; 
as such, the integrated contact intensity
establishes non-trivial connections 
between properties such as 
the energy, the pair distribution function and the momentum distribution
of the two-component
Fermi gas with zero-range interactions.
For concreteness, we 
restrict ourselves to the 
zero temperature situation, where
expectation values are calculated for a
given many-body eigenstate
$\psi(\vec{r}_1,\cdots,\vec{r}_N)$.
We note, however, that the Tan relations can be generalized
to finite temperature~\cite{tan08a}.

(i) The {\em{adiabatic energy relation}}~\cite{tan08a,tan08b,tan08c} 
defines the integrated contact intensity 
$I(a_s)$ 
through
\begin{eqnarray}
\label{eq_adiabatic}
I_{\mathrm{adia}}(a_s)= \frac{4 \pi m}{\hbar^2}  
\frac{\partial E(a_s,0)}{\partial \left(-a_s^{-1}\right)}.
\end{eqnarray}

(ii) The {\em{virial relation}}~\cite{tan08a,tan08b,tan08c} 
defines the integrated contact intensity $I(a_s)$
through
\begin{eqnarray}
\label{eq_virial}
I_{\mathrm{virial}}(a_s) = \frac{8 \pi m a_s}{\hbar^2} 
\left[  2 V_{\mathrm{tr}}(a_s,0) - E(a_s,0) \right], 
\end{eqnarray}
where
$V_{\mathrm{tr}}(a_s,r_0)$ denotes the expectation 
value of the trapping potential,
$V_{\mathrm{tr}}=\langle \sum_j m \omega^2 r_j^2/2 \rangle$.
We refer to $V_{\mathrm{tr}}$ as trap energy.

(iii) The {\em{pair relation}}~\cite{tan08a,tan08b,tan08c} defines
the integrated contact intensity in terms of the number $N_{\mathrm{pair}}^{r<s}$ of up-down pairs 
with 
distances $r$ smaller than $s$,
\begin{eqnarray}
\label{eq_pair1}
I_{\mathrm{pair}}(a_s)= \lim_{s\rightarrow 0} \frac{4 \pi \, N_{\mathrm{pair}}^{r<s}}{s}.
\end{eqnarray}
If $P_{\mathrm{pair}}(r)$ denotes the pair distribution function
for the up-down distance $r$,
normalized such that
\begin{eqnarray}
\label{eq_pairnorm}
4 \pi \int_0^{\infty} P_{\mathrm{pair}}(r) r^2 dr = N_{\uparrow} N_{\downarrow},
\end{eqnarray}
then Eq.~(\ref{eq_pair1}) can alternatively be expressed as
\begin{eqnarray}
\label{eq_pair2}
I_{\mathrm{pair}}(a_s)= \lim_{r \rightarrow 0} (4 \pi)^2 \, P_{\mathrm{pair}}(r) \, r^2 ,
\end{eqnarray}
where $P_{\mathrm{pair}}(r)$ is understood to be calculated for zero-range 
interactions.

(iv) The {\em{momentum relation}}~\cite{tan08a,tan08b,tan08c}
defines the integrated contact intensity in terms of the number 
$N_{\mathrm{atom}}^{k>K}$ of atoms with momentum $k$ greater than $K$,
\begin{eqnarray}
\label{eq_momentum1}
I_{k}(a_s) = \lim_{K \rightarrow \infty} \pi^2 \, K \, N_{\mathrm{atom}}^{k>K}.
\end{eqnarray}
This momentum relation can be reexpressed  in terms of the
momentum distribution $n_{\uparrow}(\vec{k})$ of the up-atoms and
in terms of the momentum distribution
$n_{\downarrow}(\vec{k})$  of the down atoms:

(iva) If $n_{\uparrow}(\vec{k})$ 
is normalized such that
\begin{eqnarray}
\int n_{\uparrow}(\vec{k}) d^3\vec{k}= N_{\uparrow},
\end{eqnarray}
then Eq.~(\ref{eq_momentum1})
can alternatively be expressed as
\begin{eqnarray}
\label{eq_momentum2}
I_{k,\uparrow}(a_s)=
\lim_{k^{-1} \rightarrow 0} 2 \, \pi^2 \, 4 \pi \,
\frac{n_{00,\uparrow}(k)}{\sqrt{4 \pi}} \,
k^4,
\end{eqnarray}
where $n_{00,\uparrow}(k)$
is defined through the 
partial wave expansion
\begin{eqnarray}
\label{eq_momentumexpansion}
n_{\uparrow}(\vec{k}) = \sum_{lm} n_{lm,\uparrow}(k) Y_{lm}(\hat{k})
\end{eqnarray}
and where 
$n_{\uparrow}(\vec{k})$ is 
 understood to be calculated for zero-range 
interactions. 
The partial wave decomposition
of $n_{\uparrow}(\vec{k})$ has been introduced to emphasize 
that $I_{k,\uparrow}(a_s)$ only depends on the spherically symmetric
component of $n_{\uparrow}(\vec{k})$, i.e., only on $k$ and not on $\vec{k}$.

(ivb) 
Alternatively, Eq.~(\ref{eq_momentum1}) can be rewritten in terms of the
momentum distribution $n_{\downarrow}(\vec{k})$ of the down-atoms.
To this end,
the subscript ``$\uparrow$'' in (iva) needs to replaced by ``$\downarrow$''.

For $N_{\uparrow}=N_{\downarrow}$,
we have $n_{\uparrow}(\vec{k})=n_{\downarrow}(\vec{k})=n(\vec{k})$
and $n_{00,\uparrow}({k})=n_{00,\downarrow}({k})=n_{00}({k})$, and consequently
$I_{k,\uparrow}(a_s)=I_{k,\downarrow}(a_s)=I_k(a_s)$.
For spin-imbalanced two-component
Fermi gases, in contrast,
relations (iva) and (ivb) establish a non-trivial relationship
between the momentum distributions
of the up- and down-atoms.

The quantities $I_{\mathrm{adia}}$, $I_{\mathrm{virial}}$,
$I_{\mathrm{pair}}$, $I_{k,\uparrow}$ and
$I_{k,\downarrow}$ have been predicted to 
be identical~\cite{tan08a,tan08b}. 
To the best of our knowledge this has, however, not 
yet been illustrated
explicitly. In fact, the determination of the integrated contact intensity has
so far only been pursued for a few selected systems and
limiting cases~\cite{tan08a,tan08b,tan08c,braa08,braa08a}. 
Section~\ref{sec_contactfourbody} presents explicit calculations for the 
energetically lowest-lying gas-like state of the four-body
fermion system  throughout the 
crossover and numerically verifies the equivalence of the
definitions (i) through (iv).

\section{Integrated contact intensity for four-fermion system}
\label{sec_contactfourbody}
This section determines the integrated contact intensity $I(a_s)$ 
for the energetically lowest-lying gas-like state 
of the trapped four-fermion system with $N_{\uparrow}=N_{\downarrow}=2$
as a function of the $s$-wave scattering length $a_s$ according to
definitions (i) through (iv).
Unlike for the three-fermion problem (see, e.g., Ref.~\cite{niel01}),
no analytical solutions
are known for
the four-fermion problem with zero-range interactions.
Here, we solve the stationary Schr\"odinger equation
for the four-fermion system
numerically using the correlated Gaussian (CG)
approach~\cite{cgbook,sore05,stec07c,blum07,stec08}.

The CG approach\cite{cgbook,sore05,stec07c,blum07,stec08} 
expands the wave function in terms of 
Gaussian basis functions $\Phi_l$,
\begin{eqnarray}
\label{eq_expansion}
\psi(\vec{r}_1,\cdots,\vec{r}_N)=\sum_{l=1}^{B} c_l \left[
{\cal{A}} \Phi_l(r_{12},\cdots,r_{N-1,N}) \right],
\end{eqnarray}
where the $c_k$ denote expansion coefficients and 
\begin{eqnarray}
\label{eq_gaussianbasis}
\Phi_l(r_{12},\cdots,r_{N-1,N})=\exp
\left( -\sum_{j<k}^N \frac{r_{jk}^2}{2 d_{l,jk}^2} \right).
\end{eqnarray}
Each basis function $\Phi_l$ is parameterized by $N(N-1)/2$ widths $d_{l,jk}$
(one for each interparticle distance), which
are optimized semi-stochastically following the ideas outlined in 
Ref.~\cite{cgbook}. 
In Eq.~(\ref{eq_expansion}), $B$ denotes the number of 
unsymmetrized basis functions.
For the four-fermion system with two up- and two down-atoms,
the anti-symmerization operator ${\cal{A}}$ can be conveniently
written in terms of the permutation operator $P_{jk}$, which exchanges the
position vectors of the $j$th and $k$th atom.
If the position vectors $\vec{r}_1$ and $\vec{r}_2$ belong
to
the up-atoms and the position vectors 
$\vec{r}_3$ and $\vec{r}_4$ to the down-atoms,
then
${\cal{A}}$ can be written as
\begin{eqnarray}
{\cal{A}} = 1-P_{12}-P_{34}+P_{12}P_{34}.
\end{eqnarray}
The basis functions $\Phi_l$ have vanishing total angular momentum $L$
and positive parity $\pi$ and are thus well suited to describe 
the energetically lowest-lying gas-like
state of the four-fermion
system which has $L^{\pi}=0^+$ symmetry~\cite{stec07c,blum07,stec08} .

We parametrize the two-body potential between each pair of up-down atoms
by a simple Gaussian potential with range $r_0$ and depth $V_0$,
\begin{eqnarray}
V_{\mathrm{tb}}(r)= -V_0 \exp \left( - \frac{r^2}{2 r_0^2} \right).
\end{eqnarray}
For this interaction potential, 
the Hamiltonian matrix elements 
can be calculated analytically~\cite{cgbook}.
Throughout,
we restrict ourselves to two-body potentials
that support either no or one 
$s$-wave two-body bound state in free-space.
For a given range
$r_0$, the depth $V_0$ ($V_0 \ge 0$) is adjusted 
so that the scattering length $a_s$ takes the desired value. 
While $V_{\mathrm{tb}}$ leads, in general, to non-vanishing higer partial
wave scattering lengths for finite $r_0$, the importance of these higher 
partial wave contributions
decreases with decreasing $r_0$ and vanishes
for $r_0=0$.
To mimick the zero-range limit, we perform calculations for various $r_0$
while keeping the $s$-wave scattering length fixed and extrapolate 
the quantity of interest to $r_0 =0$ (see below).
For the parameter combinations considered in this paper,
we find that the ground state of the four-fermion system has gas-like
character, i.e., that self-bound trimer and tetramer states are absent,
in agreement with findings for zero-range 
interactions~\cite{petr03,petr04aa}.

The CG approach results in an upper bound to the
exact ground state energy. Our calculations reported below 
employ between $B=450$ and 500 basis functions; 
we have checked that larger basis sets do not lead to 
a significant reduction of the energy. 
Once the expansion coefficients $c_l$ ($l=1,\cdots,B$)
have been determined by diagonalizing the generalized eigenvalue
problem (the linear dependence of the
basis functions gives rise to a non-diagonal overlap matrix),
we calculate other expectation values such as $V_{\mathrm{tr}}(a_s,r_0)$, $P_{\mathrm{pair}}(r)$
and $n_{00}(k)$.
The matrix elements associated with these observables can be determined
analytically and are readily implemented.
The computational time
for the determination of the ground state energy and the structural
properties for a given $a_s$ and $r_0$ amounts to about an hour on 
a single processor of a state of the art 
desktop computer.

Our calculations cover the weakly-attractive and weakly-repulsive  
regimes as well as the strongly-interacting unitary regime;
in particular, we consider 41 scattering lengths in the interval
$a_{\mathrm{ho}}/a_s \in [-10,10]$.
For negative scattering lengths $a_s$, we find that both 
the total energy $E(a_s,r_0)$
and the trap energy $V_{\mathrm{tr}}(a_s,r_0)$ vary linearly with
$r_0$ for fixed $a_s$.
For negative $a_s$, we typically consider three different ranges $r_0$,
i.e., $r_0=0.01, 0.03$ and $0.05a_{\mathrm{ho}}$,
and determine the zero-range quantities
$E(a_s,0)$ and $V_{\mathrm{tr}}(a_s,0)$ by
performing a linear two-parameter fit.
We checked for selected scattering lengths that the inclusion of additional
ranges leaves the fitting parameters essentially unchanged.
As the inverse scattering length $a_s^{-1}$ is increased to positive values,
the total energy starts depending non-linearly on the range $r_0$.
In this regime, we consider six different 
ranges $r_0$ ($r_0 \le 0.03a_{\mathrm{ho}}$)
for each $a_s$ and determine the zero-range limit 
$E(a_s,0)$ by performing
a quadratic three-parameter fit.
We find that a three-parameter fit provides
a reliable description of the trap energy
$V_{\mathrm{tr}}(a_s,r_0)$ for $a_{\mathrm{ho}}/a_s =0 $ to 2. For larger
inverse scattering lengths, $V_{\mathrm{tr}}(a_s,r_0)$ 
is significantly smaller than 
the absolute value of the total energy and has thus a comparatively
large uncertainty.
For $a_{\mathrm{ho}}/a_s>2$, we determine the zero-range limit
of $V_{\mathrm{tr}}(a_s,r_0)$ by performing a linear two-parameter fit.
While $V_{\mathrm{tr}}(a_s,r_0)$
may depend non-linearly on $r_0$ in this regime, our
CG data do not allow for a reliable determination of the
functional dependence.

\begin{figure}
\vspace*{+1.5cm}
\includegraphics[angle=0,width=70mm]{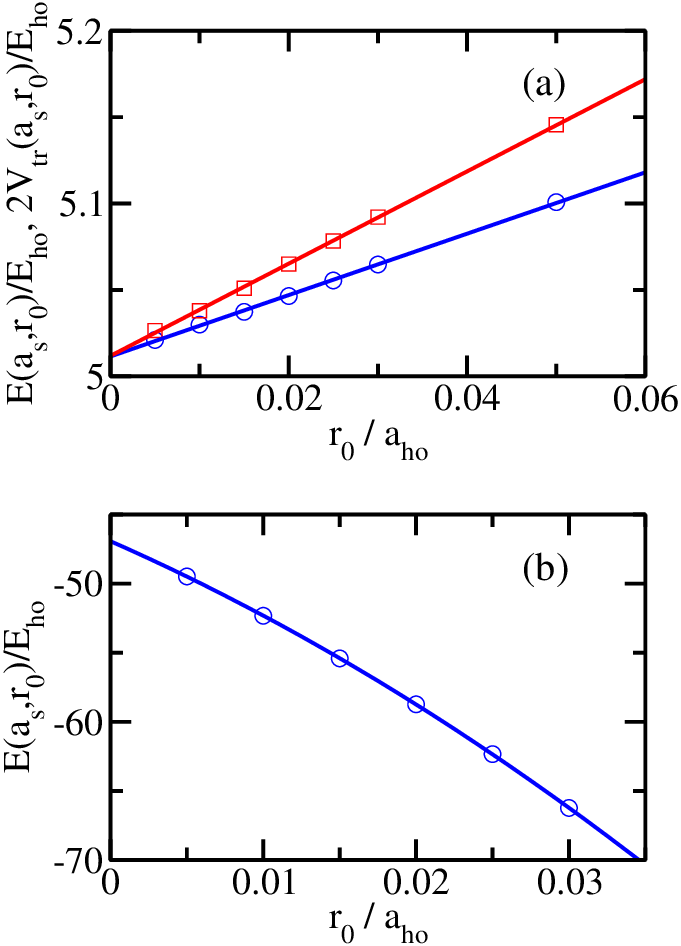}
\vspace*{0.1cm}
\caption{(Color online)
(a)
Circles and squares show $E(a_s,r_0)$ and $2 V_{\mathrm{tr}}(a_s,r_0)$,
calculated using the CG approach, 
as a function of the range $r_0$ of the two-body
interaction potential at unitarity, i.e., for $a_{\mathrm{ho}}/a_s = 0$.
Solid lines show linear fits to the CG data.
(b) 
Circles show $E(a_s,r_0)$,
calculated using the CG approach, 
as a function of $r_0$ for $a_{\mathrm{ho}}/a_s = 5$.
The solid line shows a quadratic fit to the CG energies.
For both panels,
the basis set error of the CG
data is estimated to be  smaller than the symbol
size.
}\label{fig_energyrange}
\end{figure}

As an example, 
circles and squares in Fig.~\ref{fig_energyrange}(a)
show $E(a_s,r_0)$ and $2 V_{\mathrm{tr}}(a_s,r_0)$
in units
of the oscillator energy $E_{\mathrm{ho}}$, $E_{\mathrm{ho}}=\hbar \omega$,
obtained by the CG approach as a function of the range 
$r_0$ at unitarity. Figure~\ref{fig_energyrange}(a)
shows that
the CG quantities are well described by a linear fit
(solid lines).
Furthermore,
the four-body results shown in Fig.~\ref{fig_energyrange}(a)
demonstrate the validity of the 
virial theorem for zero-range
interactions at unitarity~\cite{thom05,wern06,son07,mehe07}, 
which states that
the energy equals 
twice the trap energy. We find that the linear fit to
$E(\infty,r_0)$ and
$2V_{\mathrm{tr}}(\infty,r_0)$, which includes
all seven data points shown in Fig.~\ref{fig_energyrange}(a), 
predicts the same intercepts within errorbars.
In particular, we find $E(\infty,0)= 5.0115(5) \hbar \omega$ and
$2V_{\mathrm{tr}}(\infty,0)=5.0118(6) \hbar \omega$,
where
the number in round brackets indicates the uncertainty
that results from the fit alone (excluding possible basis set errors
of $E$ and $V_{\mathrm{tr}}$ themselves). Our energy
$E(\infty,0)$ is slightly higher than the value of $5.0096 \hbar
\omega$ of Ref.~\cite{stec09}; 
this slight difference can be attributed to a tiny
basis set error.

To illustrate the comparatively
strong dependence of $E(a_s,r_0)$ on $r_0$ for positive
scattering lengths,
Fig.~\ref{fig_energyrange}(b) shows the total energy
as a function of $r_0$ for
$a_{\mathrm{ho}}/a_s=5$.
In this case, the three-parameter fit results in a
zero-range energy $E(a_s=a_{\mathrm{ho}}/5,0)$  of $-46.93(3) \hbar \omega$, 
where
the number in round brackets indicates the uncertainty
that results from the fit alone (excluding possible basis set errors
of the energies themselves). This uncertainty is larger than 
the uncertainty of $V_{\mathrm{tr}}(a_s,0)$; in particular, we find
$V_{\mathrm{tr}}(a_s=a_{\mathrm{ho}}/5,0)=1.619(2) \hbar \omega$.
Even though the basis set extrapolation error
associated with $V_{\mathrm{tr}}(a_s,r_0)$ might be somewhat larger
than that associated with $E(a_s,r_0)$, our analysis
 suggests that the main uncertainty of the 
integrated contact intensities $I_{\mathrm{adia}}(a_s)$ 
and $I_{\mathrm{virial}}(a_s)$ 
in the small positive scattering length regime 
originates in the difficulty
of determining the zero-range limit of $E(a_s,r_0)$
more reliably.

\begin{figure}
\vspace*{+1.5cm}
\includegraphics[angle=0,width=70mm]{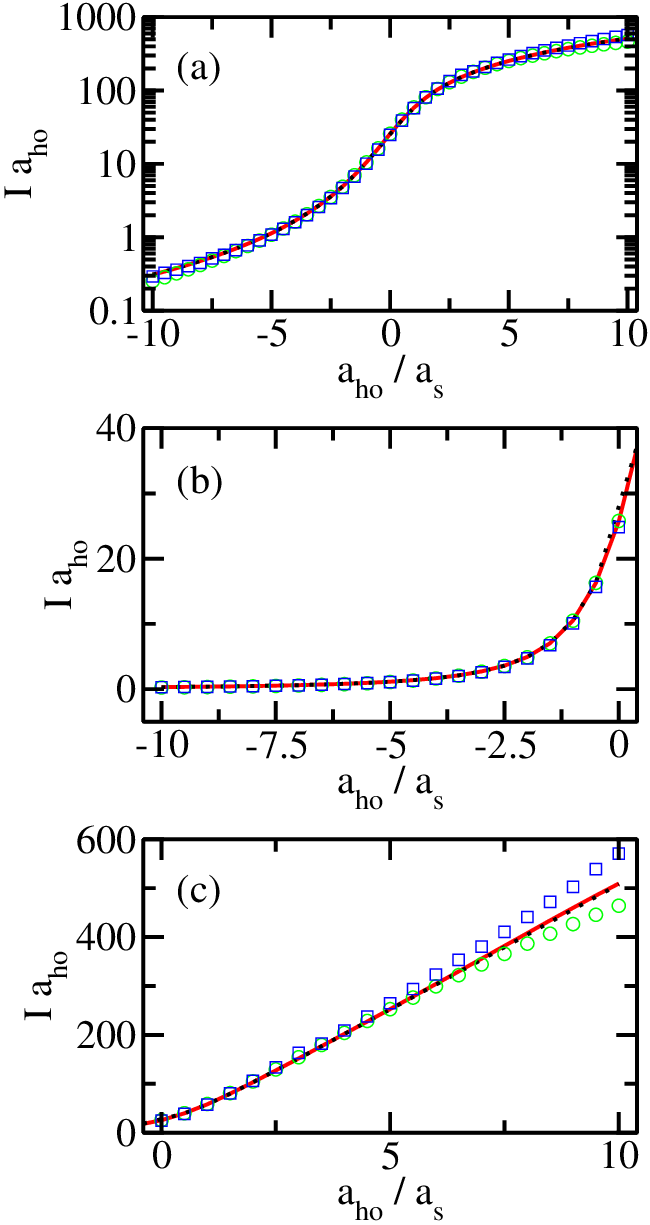}
\vspace*{0.1cm}
\caption{(Color online)
Integrated contact intensities $I_{\mathrm{adia}}(a_s)$, 
$I_{\mathrm{virial}}(a_s)$, 
$I_{\mathrm{pair}}(a_s)$
and $I_k(a_s)$
as a function of the inverse 
$s$-wave scattering length $a_s^{-1}$.
Solid and dotted lines show the integrated contact intensities
$I_{\mathrm{adia}}(a_s)$ and $I_{\mathrm{virial}}(a_s)$
calculated according to Eqs.~(\protect\ref{eq_adiabatic}) and 
(\protect\ref{eq_virial}),
respectively, using the extrapolated zero-range quantities
as input
(the two data sets are nearly indistinguishable on the scale shown and
compared in more detail in Fig.~\protect\ref{fig_contactdifference}).
Circles and squares show the integrated contact intensities
$I_{\mathrm{pair}}(a_s)$ and $I_k(a_s)$, respectively, determined 
from the pair distribution functions and the momentum distributions
for $r_0=0.005a_{\mathrm{ho}}$ (see text for details).
Panel~(a) shows the entire crossover region on a logscale, while
panels~(b) and (c) show the negative and positive scattering length regions
on a linear scale.
}\label{fig_contact}
\end{figure}

The extrapolated zero-range quantities $E(a_s,0)$ and $V_{\mathrm{tr}}(a_s,0)$
determine the integrated contact intensities 
$I_{\mathrm{adia}}(a_s)$ and $I_{\mathrm{virial}}(a_s)$ [see
definitions (i) and (ii) above].
To determine $I_{\mathrm{adia}}(a_s)$, 
we interpolate $E(a_s,0)$ and calculate its derivative 
with respect to $a_s^{-1}$ based
on this interpolation. 
Solid lines in Figs.~\ref{fig_contact}(a)-\ref{fig_contact}(c) 
show the integrated
contact intensity $I_{\mathrm{adia}}(a_s)$ 
as a function of the inverse scattering length $a_s^{-1}$.
Figure~\ref{fig_contact} shows that $I_{\mathrm{adia}}(a_s)$ 
increases monotonically with increasing
$a_s^{-1}$. This monotonic increase reflects the increase of the
two-body attraction with increasing $a_s^{-1}$. 
In going from the weakly-attractive regime ($a_{\mathrm{ho}}/a_s\le -10$)
to the weakly-repulsive regime ($a_{\mathrm{ho}}/a_s \ge 10$), 
the integrated contact intensity changes by about three orders of magnitude.
For comparison, dotted lines show the integrated contact intensity $I_{\mathrm{virial}}(a_s)$.
The two contact intensities 
$I_{\mathrm{adia}}(a_s)$ and $I_{\mathrm{virial}}(a_s)$
are nearly indistinguishable on the scales shown in
Figs.~\ref{fig_contact}(a)-\ref{fig_contact}(c), thereby lending
strong support for the equivalence of definitions (i) and (ii).

\begin{figure}
\vspace*{+1.5cm}
\includegraphics[angle=0,width=70mm]{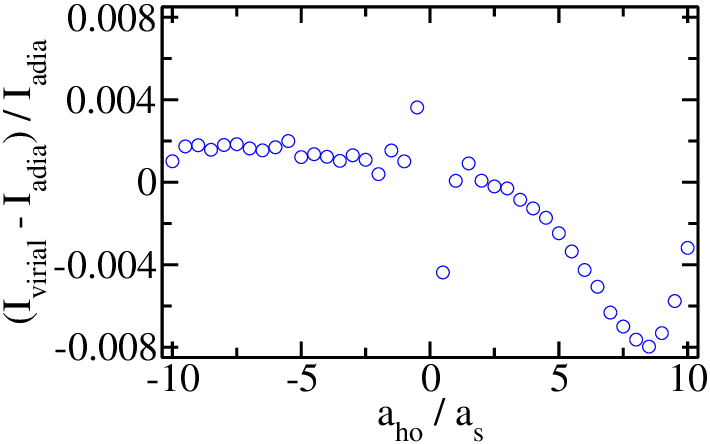}
\vspace*{0.1cm}
\caption{(Color online)
Fractional difference $[I_{\mathrm{virial}}(a_s)-I_{\mathrm{adia}}(a_s)]/I_{\mathrm{adia}}(a_s)$
as a function of the inverse $s$-wave scattering length
$a_s^{-1}$. 
}\label{fig_contactdifference}
\end{figure}

To quantify the agreement of $I_{\mathrm{adia}}(a_s)$ and $I_{\mathrm{virial}}(a_s)$,
Fig.~\ref{fig_contactdifference}
shows the fractional difference between these two quantities.
Figure~\ref{fig_contactdifference} shows that
the fractional difference is very small 
(less than 0.004) 
for negative scattering lengths
where our extrapolated zero-range quantities $E(a_s,0)$
and $V_{\mathrm{tr}}(a_s,0)$ have the smallest uncertainties.
For positive scattering lengths, the 
fractional difference shows a fairly systematic deviation
from zero that might be
attributed to small systematic errors
of $E(a_s,0)$ and $V_{\mathrm{tr}}(a_s,0)$.
The better than 1\% agreement between $I_{\mathrm{adia}}(a_s)$
and $I_{\mathrm{virial}}(a_s)$ over  the entire scattering length range considered
lends
strong numerical support for the prediction
that Eqs.~(\ref{eq_adiabatic}) and (\ref{eq_virial})
constitute equivalent definitions of the integrated contact intensity $I(a_s)$.

\begin{figure}
\vspace*{+1.5cm}
\includegraphics[angle=0,width=70mm]{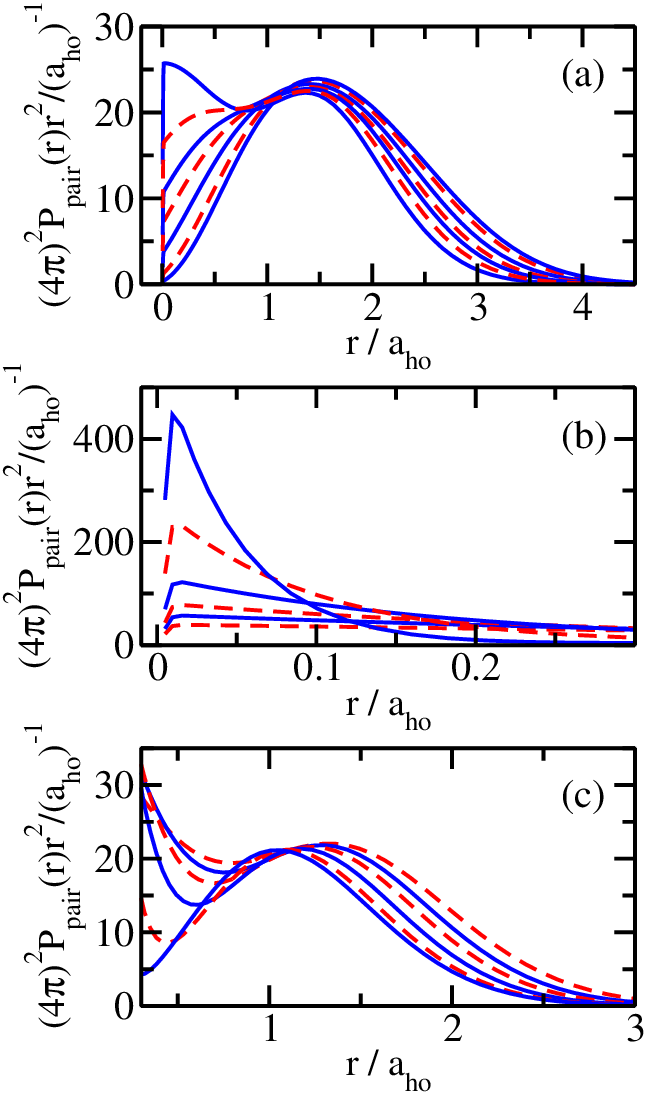}
\vspace*{0.1cm}
\caption{(Color online)
Scaled pair distribution functions $(4 \pi)^2 P_{\mathrm{pair}}(r)r^2$
as a function of $r$ for $r_0=0.005a_{\mathrm{ho}}$.
Panel~(a) shows the scaled pair distribution functions for
$a_{\mathrm{ho}}/a_s=-10$, $-5$, $-2.5$, $-1.5$, $-1$, $-0.5$ and
$0$ (from bottom to top at $r=0.25a_{\mathrm{ho}}$). 
Panels~(b) and (c) 
show the scaled pair distribution functions
for $r \le 0.3a_{\mathrm{ho}}$ and $r \ge 0.3a_{\mathrm{ho}}$, respectively,
for $a_{\mathrm{ho}}/a_s=0.5$, $1$, $1.5$, $2.5$, $5$ and $10$
[from (b) bottom to top at $r=0.05a_{\mathrm{ho}}$ 
and (c) top to bottom at $r=2a_{\mathrm{ho}}$].
Note the different scales of the axes in panels~(a) through (c).
}\label{fig_pair}
\end{figure}

Next, we discuss the determination of the integrated contact intensity
from the pair distribution functions $P_{\mathrm{pair}}(r)$
for the up-down distance.
Figure~\ref{fig_pair} shows examplary scaled
pair distribution functions
for $r_0=0.005a_{\mathrm{ho}}$
and various $s$-wave scattering lengths $a_s$ (see also Ref.~\cite{stec08}).
In particular, Fig.~\ref{fig_pair}(a) shows 
$(4 \pi)^2 P_{\mathrm{pair}}(r)r^2$
for negative $a_s$ while Figs.~\ref{fig_pair}(b) and 
\ref{fig_pair}(c) show
respectively
the small $r$ and the large $r$ regions of
$(4 \pi)^2 P_{\mathrm{pair}}(r)r^2$ for positive $a_s$.
The overall behavior of the pair distribution functions
can be described as follows:
(i) The amplitude at large $r$ decreases with increasing $a_s^{-1}$, 
reflecting the fact that the four-fermion system 
becomes more compact with increasing attraction.
(ii) As $a_s^{-1}$ increases from $-10a_{\mathrm{ho}}^{-1}$ to 0
(and to even larger values), the scaled pair distribution functions
develop a two peak structure that reflects the formation of pairs. 
The peak at smaller interparticle up-down distances $r$
corresponds to the formation of a pair while the peak at larger $r$
values ($r \approx 1a_{\mathrm{ho}}$ to $1.5a_{\mathrm{ho}}$)
reflects the fact that the second up-atom and the second down-atom are
pushed away from the first pair due to the 
Pauli exclusion principle (i.e., the formation of
self-bound trimers and tetramers is prohibited).
(iii) The scaled pair distribution functions vanish at $r=0$
for all
$s$-wave scattering lengths;
for numerical reasons, the first point of $P_{\mathrm{pair}}(r)$
is not calculated at $r=0$ but at $r=0.005a_{\mathrm{ho}}$. 
The vanishing of $P_{\mathrm{pair}}(r)r^2$
at $r=0$ is accompanied by a sharp drop 
of $P_{\mathrm{pair}}(r)r^2$
at $r$ values of the order of a few
times the range $r_0$. 
The
scaled pair distribution functions behave universally
when $r$ is larger than a few times the range $r_0$;
for smaller $r$,
the pair distribution functions
acquire
non-universal behavior.

\begin{figure}
\vspace*{+1.5cm}
\includegraphics[angle=0,width=70mm]{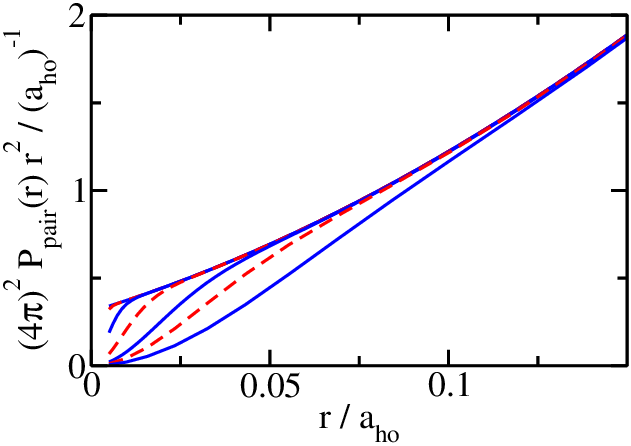}
\vspace*{0.1cm}
\caption{(Color online)
Scaled pair distribution functions $(4 \pi)^2 P_{\mathrm{pair}}(r)r^2$
as a function of $r$, $r \in [0,0.15a_{\mathrm{ho}}]$,
 for $a_{\mathrm{ho}}/a_s=-10$ and various $r_0$.
At small $r$, the scaled pair distribution functions correspond 
from bottom to top to
$r_0=0.05$, $0.03$, $0.02$, $0.01$, $0.005$, $0.0025$, and $0.001a_{\mathrm{ho}}$.
Note that the solid and dashed
lines for the two smallest ranges, i.e., for $r_0=0.001$ and
$0.0025a_{\mathrm{ho}}$, are 
nearly indistinguishable on the scale shown.
}\label{fig_pairkevin}
\end{figure}

To illustrate the universal, range independent behavior of 
the scaled pair distribution functions,
Fig.~\ref{fig_pairkevin} shows 
the quantity $(4 \pi)^2 P_{\mathrm{pair}}(r)r^2$
for a number of different  $r_0$ but
fixed $s$-wave scattering length, i.e., for $a_{\mathrm{ho}}/a_s = -10$
(corresponding to $a_s=-0.1 a_{\mathrm{ho}}$).
For this scattering length, 
the condition $r_0 \ll |a_s|$ is approximately fullfilled
if $r_0 \lesssim 0.02a_{\mathrm{ho}}$.
In agreement with this condition, 
the universal part of the pair distribution functions
$P_{\mathrm{pair}}(r)$, i.e., the part where $P_{\mathrm{pair}}(r)$ 
is independent of $r_0$, extends to smaller $r$ values with 
decreasing $r_0$. 
Figure~\ref{fig_pairkevin} shows that
the scaled pair distribution functions
for  small $r_0$, $r_0 \lesssim 0.02 a_{\mathrm{ho}}$,
behave approximately linearly in the regime 
$3 r_0 \lesssim r \lesssim 0.05-0.1a_{\mathrm{ho}}$;
in this regime, essentially no dependence on $r_0$ is visible.
The behavior of $P_{\mathrm{pair}}(r)r^2$, illustrated
exemplarily for $a_s=-0.1a_{\mathrm{ho}}$ in Fig.~\ref{fig_pairkevin}, suggests 
that the integrated contact intensity $I_{\mathrm{pair}}(a_s)$ 
can be determined readily 
by employing a linear two
parameter fit to the small $r$-regime where the
universal part of $(4\pi)^2 P_{\mathrm{pair}}(r) r^2$ varies linearly and where 
$P_{\mathrm{pair}}(r)$ is
calculated for a finite but sufficiently small $r_0$.
Alternatively, one might consider extrapolating
the pair distribution functions themselves to $r_0=0$;
this approach is not pursued in 
this work.

\begin{figure}
\vspace*{+1.5cm}
\includegraphics[angle=0,width=70mm]{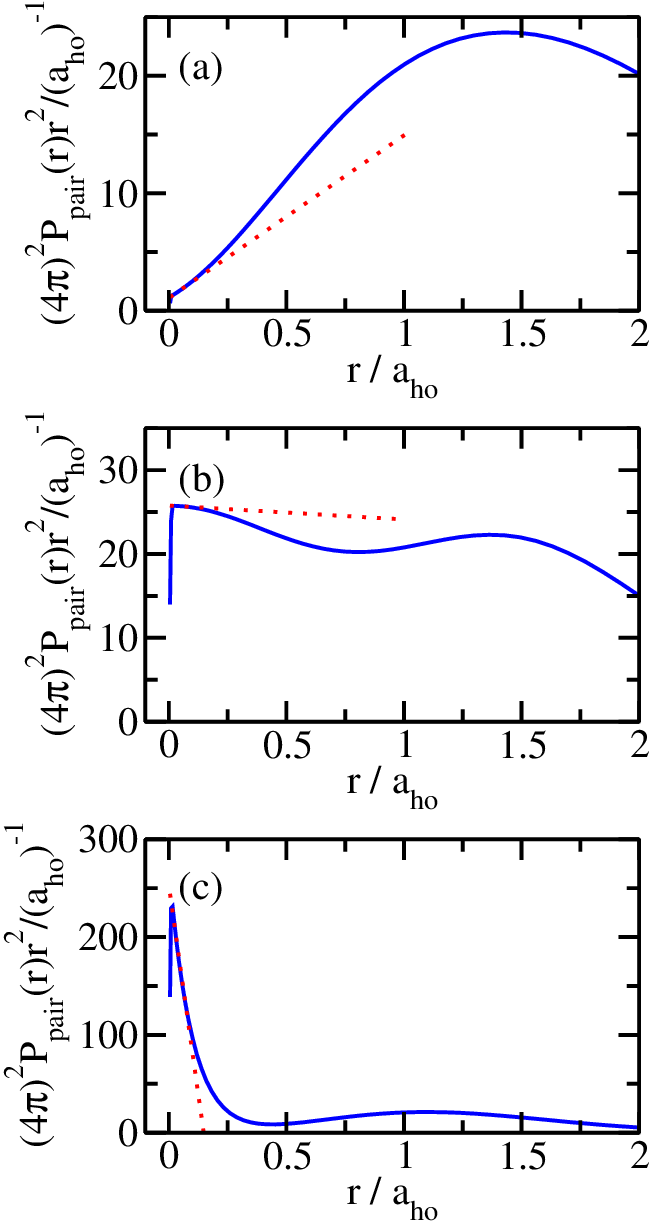}
\vspace*{0.1cm}
\caption{(Color online)
Solid lines show the
scaled pair distribution functions $(4 \pi)^2 P_{\mathrm{pair}}(r)r^2$
as a function of $r$ for $r_0=0.005a_{\mathrm{ho}}$ and 
(a) $a_{\mathrm{ho}}/a_s=-5$,
(b) $a_{\mathrm{ho}}/a_s=0$ and
(c) $a_{\mathrm{ho}}/a_s=5$.
The dotted lines are obtained by performing a linear two-parameter fit
to the small $r$-region
of the scaled pair distribution
functions; the fit includes $r$-values in the 
range  $[0.015a_{\mathrm{ho}},0.1a_{\mathrm{ho}}]$.
}\label{fig_contactpair}
\end{figure}

Figures~\ref{fig_contactpair}(a)-\ref{fig_contactpair}(c)
illustrate our determination of $I_{\mathrm{pair}}(a_s)$ for $a_{\mathrm{ho}}/a_s=-5$, 0 
and 5.
We find that $I_{\mathrm{pair}}(a_s)$, which is determined by the $r=0$ value of the fit
(i.e., the intercept),
can be determined most reliably 
for large $|a_s|$ where the universal part of $P_{\mathrm{pair}}(r)r^2$
that varies linearly extends over a comparatively large $r$ range and where 
the slope is comparatively shallow.
For small $|a_s|$, the extrapolated
integrated contact intensity $I_{\mathrm{pair}}(a_s)$ shows
a notable dependence on the fitting range employed. 
The integrated contact intensity $I_{\mathrm{pair}}(a_s)$, determined from the pair
distribution functions
for $r_0=0.005a_{\mathrm{ho}}$, 
is shown by circles
in Fig.~\ref{fig_contact}. While $I_{\mathrm{pair}}(a_s)$ compares very
favorably with $I_{\mathrm{adia}}(a_s)$ and $I_{\mathrm{virial}}(a_s)$ 
in the strongly-interacting
regime, $I_{\mathrm{pair}}(a_s)$ is notably smaller than 
$I_{\mathrm{adia}}(a_s)$ and $I_{\mathrm{virial}}(a_s)$
in the weakly-attractive
and weakly-repulsive 
regimes.
It is clear from Figs.~\ref{fig_contactpair}(a)
through \ref{fig_contactpair}(c) that 
$I_{\mathrm{pair}}(a_s)$ would be somewhat
larger in the weakly-interacting regimes if the upper fitting 
limit was reduced somewhat
(see also the caption of Fig.~\ref{fig_contactpair}). 
This would bring $I_{\mathrm{pair}}(a_s)$ into even better 
agreement with $I_{\mathrm{adia}}(a_s)$
and $I_{\mathrm{virial}}(a_s)$ in the weakly-interacting regimes while
leaving the behavior in the strongly-interacting regime
essentially unchanged.

\begin{figure}
\vspace*{+1.5cm}
\includegraphics[angle=0,width=70mm]{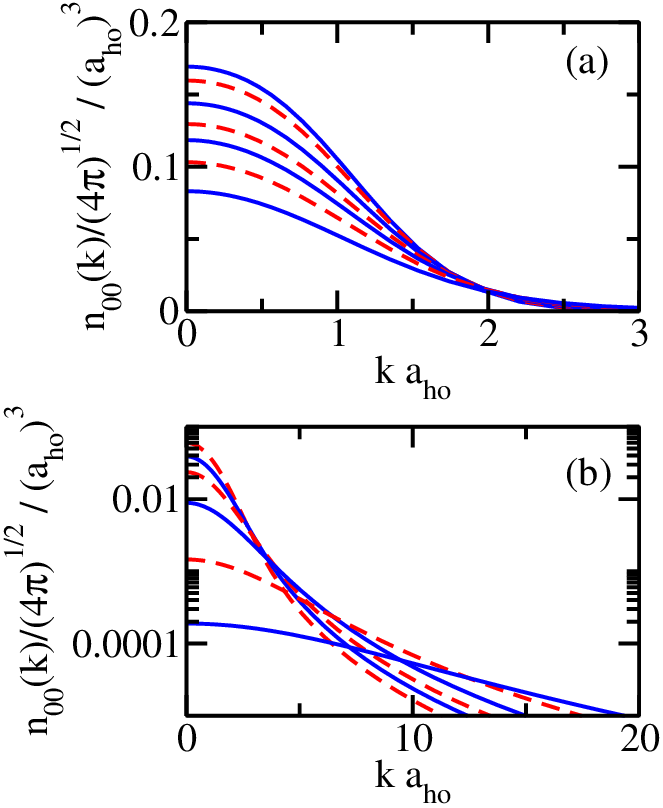}
\vspace*{0.1cm}
\caption{(Color online)
Scaled projected momentum distribution $n_{00}(k)/\sqrt{4 \pi}$
as a function of $k$ for $r_0=0.005a_{\mathrm{ho}}$.
Panel~(a) shows the scaled projected momentum distributions for
$a_{\mathrm{ho}}/a_s=-10$, $-5$, $-2.5$, $-1.5$, $-1$, $-0.5$ and
$0$ (from top to bottom at $k=0.5a_{\mathrm{ho}}^{-1}$).
Panel~(b)
shows the scaled projected momentum distributions
for $a_{\mathrm{ho}}/a_s=0.5$, $1$, $1.5$, $2.5$, $5$ and $10$
(from top to bottom at $k=1a_{\mathrm{ho}}^{-1}$).
Note the different axes in panels~(a) and (b).
}\label{fig_momentum}
\end{figure}

Figures~\ref{fig_momentum}(a) and \ref{fig_momentum}(b)
show the lowest partial wave projection $n_{00}(k)$
of the momentum distribution for fixed $r_0$, i.e., for
$r_0=0.005a_{\mathrm{ho}}$, and
selected negative and positive scattering lengths.
For weak attraction, the
lowest partial wave projection $n_{00}(k)$ is approximately Gaussian.
As $a_s^{-1}$ increases, $n_{00}(k)$ develops a tail 
at large $k$ while the small $k$ amplitude decreases.
The  tail at large $k$ indicates that the system is characterized by 
increasingly small length scales, reflecting the formation of pairs.

\begin{figure}
\vspace*{+1.5cm}
\includegraphics[angle=0,width=70mm]{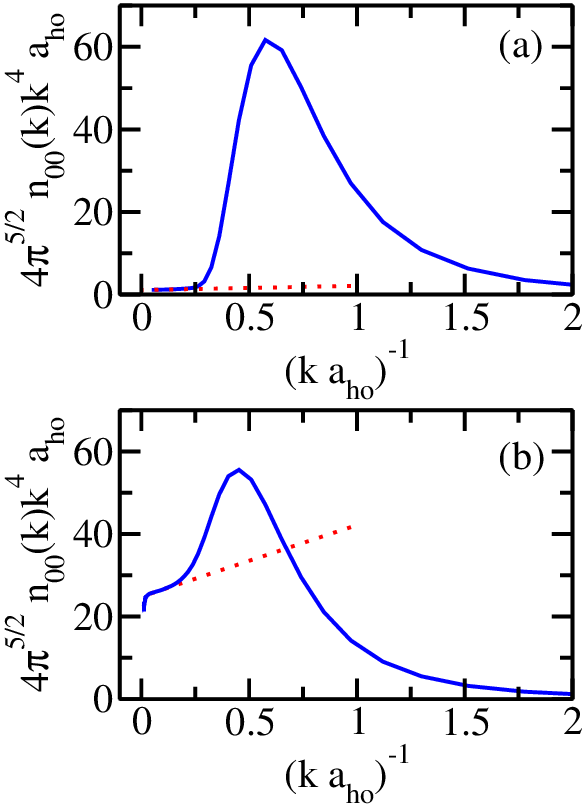}
\vspace*{0.1cm}
\caption{(Color online)
The solid lines show the scaled projected momentum distribution 
$2\pi^2 \, 4\pi \, n_{00}(k)/\sqrt{4 \pi}$
as a function of $k^{-1}$ for $r_0=0.005a_{\mathrm{ho}}$
and (a) $a_{\mathrm{ho}}/a_s=-5$ and
(b) $a_{\mathrm{ho}}/a_s=0$.
The dotted lines are obtained by performing a linear two-parameter fit
to the small $k^{-1}$-region
of the scaled  projected momentum distribution; 
the fit includes $k^{-1}$-values in the 
ranges  (a) $[0.05a_{\mathrm{ho}},0.1a_{\mathrm{ho}}]$ and
(b) $[0.03a_{\mathrm{ho}},0.1a_{\mathrm{ho}}]$.
}\label{fig_contactmomentum}
\end{figure}

To determine the integrated contact intensity 
$I_k(a_s)$ from the lowest partial wave projection
$n_{00}(k)$, we scale $n_{00}(k)$ by $k^4$ and plot the resulting quantity as a
function of $k^{-1}$ [see 
Eq.~(\ref{eq_momentum2}) and Figs.~\ref{fig_contactmomentum}(a) and 
\ref{fig_contactmomentum}(b)].
Similarly to our analysis of the scaled pair distribution functions,
we extract the integrated contact intensity $I_{k}(a_s)$
by employing a linear two-parameter fit 
to the linear small $k^{-1}$ region of
$2\pi^2 4 \pi n_{00}(k)/\sqrt{4 \pi}$, where $n_{00}(k)$ is calculated
for a small but finite $r_0$.
Figures~\ref{fig_contactmomentum}(a)
and \ref{fig_contactmomentum}(b) 
illustrate the procedure for $a_{\mathrm{ho}}/a_s=-5$ and $0$,
respectively.
The resulting
integrated contact intensities $I_k(a_s)$, determined
from the momentum distributions for $r_0=0.005a_{\mathrm{ho}}$,
are shown 
as a function of the inverse $s$-wave scattering length $a_s^{-1}$
by squares in Fig.~\ref{fig_contact}.
The agreement between $I_k(a_s)$ and the integrated contact intensities
determined based on definitions (i) and (ii) is very good for 
$a_s^{-1} \lesssim 5$. For larger inverse scattering lengths,
notable deviations are visible. These deviations can, as in the case of
$I_{\mathrm{pair}}(a_s)$, be traced back to the fitting range
employed to extract $I_k(a_s)$. Despite these deviations,
Figs.~\ref{fig_contact}(a)-\ref{fig_contact}(c) 
convincingly illustrate the equivalence of definitions (i) through (iv)
of the integrated contact intensity.

\section{Generalized virial theorem}
\label{sec_generalvirial}
Our total energies $E(a_s,r_0)$ and trap energies $V_{\mathrm{tr}}(a_s,r_0)$
for the energetically lowest-lying gas-like state of the four-fermion
system
can be readily combined to verify the generalized virial theorem
derived by Werner~\cite{wern08}. This
theorem applies not only to infinitely large $s$-wave scattering lengths
but also to finite scattering lengths and 
accounts for finite range corrections,
\begin{eqnarray}
\label{eq_werner}
E(a_s,r_0)= \nonumber \\
2 V_{\mathrm{tr}}(a_s,r_0) - 
\frac{1}{2} a_s \frac{\partial E(a_s,r_0)}{\partial a_s} -
\frac{1}{2} r_0 \frac{\partial E(a_s,r_0)}{\partial r_0}.
\end{eqnarray}
Here, it is assumed that the underlying two-body potential $V_{\mathrm{tb}}$
depends on only two lengths, the $s$-wave scattering length 
$a_s$ and the range $r_0$.
The second term on the right hand side of Eq.~(\ref{eq_werner})
can be rewritten as $a_s^{-1} [\partial E(a_s,r_0)/\partial a_s^{-1}] /2$,
which shows that it vanishes at unitarity. In the zero-range limit 
(i.e., for $r_0=0$),
Eq.~(\ref{eq_werner}) thus reduces to the ``usual'' virial
theorem $E(a_s,0)=2V_{\mathrm{tr}}(a_s,0)$~\cite{thom05,wern06,son07,mehe07},
which has been discussed in the context 
of Fig.~\ref{fig_energyrange}(a).

\begin{figure}
\vspace*{+1.5cm}
\includegraphics[angle=0,width=70mm]{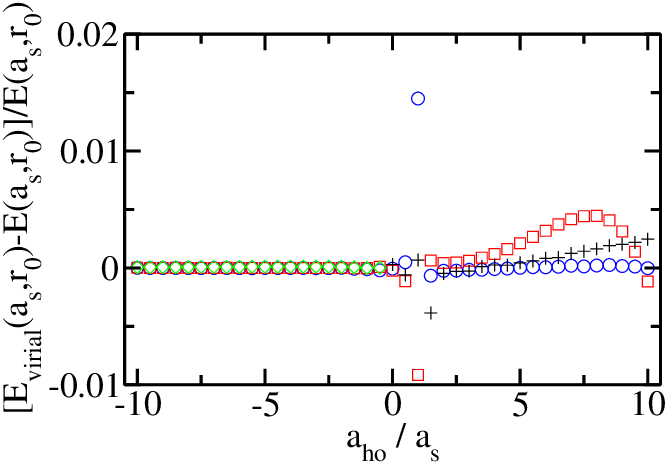}
\vspace*{0.1cm}
\caption{(Color online)
Fractional difference $[E_{\mathrm{virial}}(a_s,r_0) - E(a_s,r_0)]/E(a_s,r_0)$
as a function of the inverse $s$-wave scattering length $a_s^{-1}$.
Pluses, circles, squares and diamonds show the fractional
difference for $r_0=0.005,0.01,0.03$ and $0.05a_{\mathrm{ho}}$
(for negative $a_s$, only values for the largest three $r_0$ are shown;
for positive $a_s$, only values for the smallest three $r_0$
are shown).
For negative $a_s$, the absolute value
of the fractional difference is smaller than
$0.00025$ for all $r_0$ shown. 
}\label{fig_virialwerner}
\end{figure}

Symbols in Fig.~\ref{fig_virialwerner} 
show the fractional difference between the
four-fermion energy $E_{\mathrm{virial}}(a_s,r_0)$
calculated according to the right hand side of Eq.~(\ref{eq_werner})
and the energy calculated by the CG approach
as a function of the inverse scattering length $a_s^{-1}$.
For negative scattering lengths, the virial theorem and CG energies agree
to better than 0.025\% for the parameter combinations
considered, 
thus confirming the generalized virial theorem with high accuracy.
For positive scattering lengths, our results confirm the generalized
virial theorem with 2\% accuracy. 
We note that the validity of the generalized virial theorem has 
very recently also been
investigated for large two-component Fermi gases
by means of a Monte Carlo approach~\cite{rosa09}.

\section{Conclusions}
\label{sec_conclusion}
This paper investigates universal properties of 
dilute equal-mass two-component $s$-wave interacting Fermi gases.
In particular, our calculations are performed for the energetically
lowest-lying gas-like state of the trapped four-fermion system 
without spin-imbalance. 
We determined the energies and various other properties as functions
of the scattering length and the range of the underlying two-body potential.
Analyzing these quantities, we were able to explicitly demonstrate
the equivalence of four distinctly different definitions
of the integrated contact intensity $I(a_s)$~\cite{tan08a,tan08b,tan08c} 
and of a generalized
virial theorem~\cite{wern08} with high accuracy.
In addition, we performed selected calculations for the 
energetically lowest-lying gas-like  $L^{\pi}=0^{+}$ state 
of the three-fermion 
system with $(N_{\uparrow},N_{\downarrow})=(2,1)$
and the five-fermion system with
$(N_{\uparrow},N_{\downarrow})=(3,2)$. These calculations 
confirm our findings discussed in Sec.~\ref{sec_contactfourbody}
for the four-fermion system.
We have outlined in detail how the integrated 
contact intensities $I_{\mathrm{pair}}(a_s)$
and $I_k(a_s)$ can be extracted from our numerical data. While we have not 
commented on how to best measure the integrated contact intensity
experimentally, our outlined analysis provides some guidance as to the 
demands on the accuracy of 
experimental data if the integrated contact 
intensity is to be measured with high accuracy. At the same time, 
we have shown that the integrated contact intensity 
changes by multiple orders of magnitude
through the crossover, making a qualitative measurement of $I(a_s)$
appear
quite plausable.

The explicit verification of the universal relations away from unitarity
paves the way for developing the concept of the 
integrated contact intensity further. For example,
are there other properties of the many-body two-component Fermi gas
besides those already discussed~\cite{tan08a,tan08b,tan08c} 
that are related through
the integrated contact intensity? 
Can the concept of the integrated contact intensity
be extended to low-dimensional systems?
Furthermore,
can the concept of the integrated contact intensity be extended
to relate various properties of unequal-mass two-component Fermi gases
with large mass ratio or Bose gases?
For these systems, a three-body parameter
that is associated with Efimov physics~\cite{efim71,efim73} 
arises and it has been 
speculated~\cite{braa08}
that this three-body parameter could somehow be related to an
integrated 
contact intensity that, 
physically speaking, accounts for the formation of trimers.

Support by the NSF through
grant PHY-0855332
and by the ARO, 
and discussions with J. von Stecher 
on the optimization procedure of the CG method
are gratefully acknowledged.
KMD acknowledges partial support through a Millenium fellowship.


\end{document}